\documentclass[aps,prl,twocolumn,showpacs,letterpaper,amsmath,amssymb,superscriptaddress,
showkeys,floatfix,preprintnumbers]{revtex4}
\usepackage{epsfig}
\usepackage[latin9]{inputenc}
\usepackage{color}
\usepackage{hyperref}
\usepackage{graphicx}
\usepackage{sidecap}
\usepackage{hypernat}
\def\etal{{\em{et al.}}}
\begin{document}

\title{Re-examining the electronic structure of germanium: A first-principle study}

\author{C. E. Ekuma}
\altaffiliation{corresponding email: cekuma1@lsu.edu}
\affiliation{Department of Physics \& Astronomy Louisiana State University,
Baton Rouge, LA 70803, USA}
\affiliation{Center for Computation and Technology, Louisiana State University, Baton Rouge, LA 70803, USA}

\author{M. Jarrell}
\affiliation{Department of Physics \& Astronomy Louisiana State University,
Baton Rouge, LA 70803, USA}
\affiliation{Center for Computation and Technology, Louisiana State University, Baton Rouge, LA 70803, USA}

\author{J. Moreno}
\affiliation{Department of Physics \& Astronomy Louisiana State University,
Baton Rouge, LA 70803, USA}
\affiliation{Center for Computation and Technology, Louisiana State University, Baton Rouge, LA 70803, USA}

\author{D. Bagayoko}
\affiliation{Department of Physics, Southern University and A \& M 
College, Baton Rouge, LA 70813, USA}

%\date{\today}

\begin{abstract}
\noindent We report results from an efficient, ab-initio method for self-consistent 
calculations of electronic and structural 
properties of Ge. Our 
non-relativistic calculations employed a GGA-potential and LCAO-formalism. 
The distinctive feature of our computations stem from 
the use of Bagayoko-Zhao-Williams-Ekuma-Franklin method. Our results are 
in agreement with experimental ones where the latter are available. 
In particular, our theoretical, indirect band gap (E$^\mathrm{\Gamma-L}_g$) of 0.65 eV, 
at the experimental lattice constant of 
5.66 \AA{}, is in excellent agreement with experiment. Our predicted, equilibrium 
lattice constant is 5.63 \AA{}, with corresponding E$^\mathrm{\Gamma-L}_g$ of 0.65 eV and a bulk modulus of 80 GPa.
\end{abstract}

\maketitle

There has been great interest in germanium (Ge) due to the central role it plays in modern electronics. 
As a consequence of this 
interest, there has been many 
experimental \cite{PhysRevB.58.10394,Dr2000,PhysRevB.12.4405,PhysRevB.47.2130,
PhysRevB.42.7429,PhysRevB.73.035217,PhysRevB.49.16523,PhysRevB.9.600,PhysRevB.44.13791,
PhysRevLett.57.2579,Feenstra1987,PhysRevB.40.9644,PhysRevB.28.1965} 
as well as theoretical \cite{PhysRevB.56.6648,PhysRevB.64.245204,PhysRevLett.89.126401,Smith2012,Yutaka2009,PhysRevB.62.7121,
PhysRevB.40.9644} studies, 
with the latter utilizing computational techniques of varying sophistications, ranging from tight binding 
method, the empirical pseudopotential method, and density functional theory (DFT) methods to approaches that 
entirely go beyond DFT.

Generally, DFT \cite{PhysRev.136.B864,PhysRev.140.A1133,
Kresse2011,PhysRevLett.105.196403,PhysRevLett.75.1126} is the 
most widely used ab initio theory in modern computational studies 
of electronic properties of materials. Angle resolved photoemission and X-ray resonant 
emission spectra \cite{PhysRevB.58.10394,Dr2000} show that 
the upper valence bands for traditional $sp$ 
semiconductors and metals are well reproduced within the local density functional approximation (LDA) to DFT. 
However, the low-laying conduction bands are severely underestimated, leading to the well-known band gap problem. 

The Green function, G, and the screened Coulomb interaction, W, method (GW) 
has been used to accurately reproduce the band gap and other properties of Ge. However, the GW results do not 
lead to any significant changes in the dispersion of the upper valence bands \cite{PhysRevB.48.17791} as 
compared to DFT findings, 
but rather a rigid shift in their energies and a 
reduction of the calculated valence-band width \cite{PhysRevB.64.245204}. In most GW methods, the Dyson 
equation is not solved self-consistently \cite{PhysRevLett.89.126401}, resulting in 
a violation of charge conservation \cite{PhysRevB.56.3528}. From the work of \citet{PhysRevLett.89.126401}, 
it is evident (as pointed out by the authors) that even though their calculated band gap is in 
agreement with experiment, low-laying empty states near the $X$ point are severely underestimated. 

Heyd-Scuseria-Ernzerhof (HSE) hybrid functional \cite{heyd:1187,heyd:8207} calculations reported an 
indirect gap in the range 0.63 -- 0.77 eV \cite{PhysRevB.83.085201,PhysRevB.81.153203,PhysRevB.74.073101}. 
The latter authors \cite{PhysRevB.83.085201,PhysRevB.81.153203,PhysRevB.74.073101} 
did not publish any electronic energies at the high symmetry points 
of Ge. The HSE approach involves 
a range separation of the exchange energy into some fractions of 
nonlocal Hartree-Fock exchange and of DFT exchange potentials. 
There are several versions of HSE \cite{heyd:1187,heyd:8207,krukau:224106,janesko:084111,PhysRevB.81.153203}. 
The approach generally entails  a range separation parameter, $\omega$, 
which varies between 0 and $\infty$.  
There exists a value
of $\omega$ that gives the correct gap for a given system; 
this value is adjusted from one system to another \cite{Henderson2011,PhysRevB.81.153203,PhysRevB.74.073101}.  
 
The theoretical underestimations of band gaps and other energy eigenvalues have been ascribed to the inadequacies of 
DFT potentials for the description of 
the ground state electronic properties of Ge \cite{PhysRevLett.89.126401,PhysRevB.64.245204,PhysRevB.62.7121}. 
Other methods \cite{PhysRevLett.89.126401,PhysRevB.64.245204,PhysRevB.62.7121,PhysRevB.48.17791,PhysRevB.34.5390,PhysRevB.81.153203} that 
entirely go beyond DFT 
do not obtain the band gap value of Ge and related electronic structure quantities without adjustments or fitting parameters. 
Further, to the best of our knowledge, there is no rigorous theoretical results predicting (or addressing) the indirect 
band gap of Ge. This unsatisfactory situation is a key motivation for our work. A concomitant motivation is to attain a highly 
accurate, ab-initio, self-consistent computational capability that lends itself to informing and guiding the design and fabrication 
of semiconductor based devices. 

The possibility for the referenced attainment has been stated by Bagayoko and co-workers 
\cite{Bagayoko2008,Ekuma2011,ekuma:012189}, despite perceived limitations of DFT in the literature. 
They explained the fact that the derivative discontinuity of the 
exchange correlation energy has yet to be proven to be non-zero 
in semiconductors \cite{Bagayoko2008}, even though some believe it to be non-zero. 
Perdew \etal \cite{PhysRevLett.49.1691}, following a thought experiment on a 
diatomic molecule, established the existence of a derivative 
discontinuity of the exchange correlation energy. Generalizing this derivative 
discontinuity to the case of semiconductors, 
Perdew and Levy \cite{PhysRevLett.51.1884} showed that the exchange 
correlation potential may jump by the discontinuity, $\Delta_{xc}$, 
when the number of electrons in the system under study 
increases by one. In their work on this discontinuity in insulators, 
\citet{PhysRevLett.51.1888} explicitly stated that 
while they established the existence of the discontinuity, 
they could not determine whether it is zero or not in real insulators. 
Subsequent work by 
Sham and Schl\"{u}ter \cite{PhysRevB.32.3883} derived the discontinuity 
of the functional derivative of E$_{xc}$ in insulators by considering 
an increase of the number of electrons by one. Cautiously, 
these authors concluded that the discrepancy between 
calculated and measured band gap is a measure of 
the discontinuity  $\Delta_{xc}$ - given the results from several
calculations -- if the employed LDA potentials are assumed 
to be good approximations. 
The description of our method 
below indicates the strong possibility that some current 
LDA and GGA potentials may be very good approximations. 
Computational approaches may be sources of the discrepancy, as the 
perceived limitations of DFT, described in the literature, 
are far from being settled. This statement is 
supported in part by several DFT results, 
including predictions, in excellent agreement with experiment \cite{Bagayoko2008,Ekumab2011}. 
An ampler discussion of the derivative discontinuity and of effects of computational approaches is 
provided in the supplementary material.

Ge has an experimental, indirect band gap (E$^\mathrm{\Gamma-L}_g$) of 0.664 eV 
at room temperature \cite{Madelung2006}. 
Theoretical calculations using several techniques have led to band gaps of Ge in the ranges 
of -0.02 to 0.35 eV for LDA and GGA \cite{PhysRevB.64.245204,Persson2006} and of 0.51 to 0.94 eV 
for the GW method \cite{PhysRevLett.89.126401,PhysRevB.64.245204}. In most of these GW results, 
the position of the conduction band minimum ($CB_{min}$) is not at the $L$ point. 
For instance, in the work of \citet{PhysRevLett.89.126401}, 
it is at the $X$ point for their $\sum_{GW}$[$G_{GW}$], $\sum_{GW}$[$G_{LDA}$], and $\sum_{GW}$[$G_{LDA}$]+no 3$d's$ 
methods since their predicted band gap E$^\mathrm{\Gamma-X}_g$ value of 0.71, 0.49, and 0.49 eV are smaller than the 
corresponding 0.79, 0.51, 0.51 eV for E$^\mathrm{\Gamma-L}_g$. However, their PS-based + CPP method obtained correctly 
the position of $CB_{min}$. A similar trend can be seen in the work of \citet{PhysRevB.64.245204}
where E$^\mathrm{\Gamma-\Gamma}_g$ of  0.57 eV is smaller than E$^\mathrm{\Gamma-L}_g$ of 0.62 eV, 
and in cases where the ordering is 
correct, the gap is over-estimated by $\sim$ 0.3 eV. The work of \citet{PhysRevB.62.7121} found the lowest eigenvalue 
energies at the high symmetry points to be $\sim$ 1.0 eV over-estimated. The same can be said of several other GW results 
\cite{PhysRevB.34.5390,PhysRevB.48.17791} where the gap of Ge is found to be direct instead of indirect, as established by 
experiment. 

In this letter, we present a simple, robust approach based on basis set 
optimization \cite{Ekuma2011,ekuma:012189,Ekumab2011}.
We use this method to calculate the electronic and structural properties of Ge and 
compare our results to experiments. 
  
In the ground state, Ge crystallizes in the diamond structure (space group: 
O$_\mathrm{h}^7$ -- Fd\={3}m) \cite{Wyckoff1963,Hellwege1969,Hahn2005} 
with room temperature lattice constant of 5.66 \AA{} \cite{Hellwege1969}.

Our ab initio, self consistent, nonrelativistic calculations employed a linear combination 
of atomic orbitals (LCAO). We utilize the electronic structure package from the Ames Laboratory 
of the US Department of Energy (DOE), Ames, Iowa \cite{Harmon1982}. 
For the LDA computations, we used the Ceperley and Alder density functional approximation (DFT)
exchange-correlation functional \cite{Ceperley1980} as parameterized 
by Vosko-Wilk-Nusair \cite{Vosko1980}. We refer to it as the CA-VWN LDA potential. 
The GGA calculations were carried out using the 
Ceperley and Alder DFT exchange correlation contribution \cite{Ceperley1980} as 
parameterized by Perdew and Wang \cite{PhysRevB.43.8911,Perdew1992,Perdew1991}. We refer to it as the 
CA-PW GGA potential.  

\begin{figure}
%FIG. 1
\includegraphics*[width=1\columnwidth,clip=true]{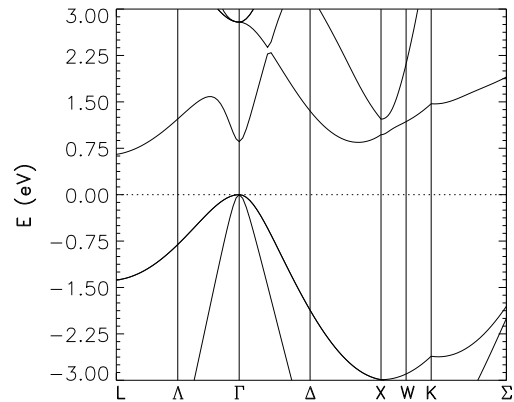}
\caption{Calculated band structure of Ge, as obtained using our GGA BZW-EF 
optimal basis set. The horizontal, dotted line denotes the position of the 
Fermi energy (E$_\mathrm{F}$) which has been set 
equal to zero. As it is evident from the plot, the $CB_{min}$ is actually 
at the $L$-point. 
The calculated indirect band gap (E$^\mathrm{\Gamma-L}_g$) of 0.65 eV is basically the same as the 
room temperature experiment one of 0.66 eV \cite{Madelung2006}.} 
\label{fig:ge_bnd}
\end{figure}

\begin{figure}
%FIG. 2
\includegraphics*[trim = 0mm 0mm 0mm 80mm,width=1\columnwidth,clip=true,angle=0]{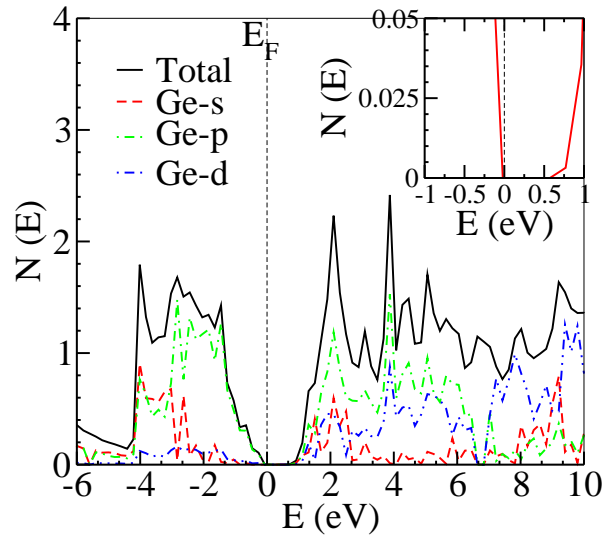}
\caption{(Color online). Calculated density of states (DOS) and partial DOS of Ge, as obtained using our GGA BZW-EF 
optimal basis set. The vertical, straight, dashed line denotes the Fermi energy (E$_\mathrm{F}$) which has been set 
equal to zero. As per the insert, it clearly shows the energy band gap.} 
\label{fig:ge_pdos}
\end{figure}

The distinctive feature of our 
calculations is the use of the Bagayoko, Zhao, Williams, Ekuma, and Franklin (BZW-EF) method 
\cite{Zhao1999, Bagayoko2007,Ekuma2010,ekuma:012189}. 
In the BZW method, \cite{Zhao1999, Bagayoko2007, Ekuma2010,ekuma:012189,Ekuma2011, Bagayoko2005, Ekumab2011, Bagayoko2004} 
a minimum of three successive, 
self-consistent calculations are performed. 
For the first one, a relatively small basis set, 
no smaller than the minimum basis set that accounts 
for all the electrons in the system, is employed. 
This set is augmented with one orbital for the second 
calculation. Depending on the $s$, $p$, $d$, or $f$ nature of 
this orbital, 2, 6, 10, and 14 functions, respectively, 
are added to the basis set. The occupied energies of 
Calculation I and II are compared graphically and numerically. 
They are generally different, with eigenvalues from Calculation 
II being lower or equal to corresponding ones from I. 
This process of adding one orbital and of performing self-consistent 
calculations is continued until a calculation is found, say N, 
to have the same occupied energies as Calculation (N+1) that 
immediately follows it. Then, the outputs of Calculation N 
provide the LDA/GGA BZW description of the material under study. 
The basis set for this calculation is the optimal basis set, i.e., 
the smallest one that leads to the minima of all the occupied energies. 
In Calculation (N+1) and others with larger basis sets that include 
the optimal basis set, the charge density, the potential, and the 
Hamiltonian do not change as compared to their values from Calculation N;  
nor do the occupied energies. These calculations, however, do lead to the 
lowering of some unoccupied energies on account of the basis 
set and variational effect stemming from the Rayleigh theorem \cite{Bagayoko1998,ekuma:012189}. 
For Ge, the basis set of Calculation III was the optimal one 
[Ge (3$s$3$p$3$d$4$s$4$p$4$d$5$p$), for the valence states only], 
where the 4$d$ and 5$p$ orbitals are unoccupied.  

The enhancement of the 
original BZW method is in the methodical increase of the basis set in our calculations. 
\cite{ekuma:032163,ekuma:012189,Franklin2013729}. This enhancement leads 
to adding $p$, $d$, and $f$ polarization orbitals, for a given principal quantum number, 
before adding the spherically symmetric $s$ orbital. These 
additional unoccupied orbitals are needed to accommodate the reorganization of the electron cloud, including 
possible polarization, in the crystal environment. For valence electrons in molecules to solids, polarization 
has primacy over spherical symmetry. The BZW method has been shown to lead to accurate ground state properties of 
many semiconductors: c-InN \cite{Bagayoko2004}, w-InN \cite{Bagayoko2005}, w-CdS 
\cite{Ekuma2011}, c-CdS \cite{ibid2011}, w-ZnO \cite{Franklin2013729}, rutile-TiO$_{2}$ \cite{Ekumab2011}, 
SrTiO$_\mathrm{3}$\cite{ekuma:012189} 
AlAs \cite{Jin2006}, GaN, Si, C, RuO$_{2}$ \cite{Zhao1999}, 
and carbon nanotubes \cite{Zhao2004}. Details of our method have been explicitly explained 
in the literatures \cite{ekuma:012189,ibid2011,Ekuma2010,Ekumab2011,Ekuma2011,Bagayoko1998,Zhao1999}.  

\begin{table*}[htb] 
\centering
%Table I
\caption{The calculated electronic energies (eV) of different bands at high symmetry points in the Brillouin 
zone of Ge, as obtained by GGA BZW-EF calculations, compared to experimental ones where the latter 
are available. Experimental data are from Refs. \cite{Hellwege1969,PhysRevB.9.3473,
PhysRevB.12.4405,PhysRevB.42.7429,PhysRevB.33.2607}. }
\begin{tabular}{cccc|cccc|cccc|ccc}
\hline\hline
\multicolumn{2}{c}{GGA BZW-EF} & EXP &  & \multicolumn{2}{c}{GGA BZW-EF} & EXP &  
& \multicolumn{2}{c}{GGA BZW-EF} & EXP &  & \multicolumn{2}{c}{GGA BZW-EF} & EXP\\
\hline
$\Gamma$ & -12.56 & -12.6 $\pm$ 0.3 &  & $K$ & -9.09 & -10.10 $\pm$ 0.2 &  & $X$ & -9.02 & 
-9.3 $\pm$ 0.2 &  & $L$ & -10.42 & -10.60 $\pm$ 0.50 \\
 & 0.00 & 0.00 &  &  & -7.98 & N/A &  &  & -8.20 & -8.65 & & & -7.36 & -7.70 $\pm$ 0.20 \\
 & 0.86 & 0.90&  &  & -4.18 & -4.20 $\pm$ 0.20 &  &  & -3.00 & -3.15 $\pm$ 0.20 & & & -1.38 & -1.40 $\pm$ 0.30 \\
 & 2.79 & 3.01 &  &  & -2.61 & N/A &  &  & 0.97 & N/A & & &  0.65 & 0.70, 0.76 \\
 & &  & &  & 1.47 & N/A &  &  &  1.22 & 1.16 & & &  4.00 & 4.20 \\
 & & & & & 4.19 & N/A & & & 9.84 & N/A & & & 8.36 & 7.90 \\
 & & & & & 7.59 & N/A & & &  \\
 & & & & & 7.81 & N/A & & &  \\
\hline
$W$ & -8.98 & N/A &  & $\sum$ & -11.55 & N/A & & $\Delta$ & -11.65 & N/A & & $\Lambda$ & -11.89 & N/A\\
 & -8.17 & N/A &  &  & -4.56 & -4.30 $\pm$ 0.20 & & & -4.14 & N/A & & & -4.19 & N/A\\
 & -3.49 & N/A & &  &  -2.00 & N/A & & & -1.87 & N/A & & & -0.81 & N/A\\
 & -2.90 & N/A &  &  & -1.80 & N/A & & & 1.35 & N/A & & & 1.22 & N/A \\
 & 1.18 & N/A & &  & 1.90 & N/A & & & 3.22 & N/A & & & 3.82 & N/A \\
& 2.11 & N/A&  &  & 3.34 & N/A & & & 5.84 & N/A & & & 5.76 & N/A\\
& 8.95 & N/A&  &  & 5.21 & N/A \\
& 9.15 & N/A&  &  & 5.56 & N/A \\
\hline\hline
\end{tabular}
\label{table:ge_energy}
\end{table*}

The Brillouin zone (BZ) integration for the charge density in the self consistent procedure is 
based on 28 special k points in the irreducible Brillouin zone (IBZ). The computational error for the valence 
charge is 2.3 x 10$^{-5}$ eV per valence electron. 
The self consistent potential converged to a difference of 10$^{-5}$ after several tens of iterations. 
The energy eigenvalues and eigenfunctions are 
then obtained at 161 special k points in the IBZ for the band structure. A total of 89 weighted k points, 
chosen along the high symmetry lines in the IBZ of Ge, are used to solve for the 
energy eigenvalues from which the electron density of states (DOS) are calculated using the 
linear analytical tetrahedron method \cite{Lehmann1972}. The partial density of states (pDOS) and the effective 
charge at each atomic sites are evaluated using the Mulliken charge analysis procedure \cite{Mulliken1955}. 
We also calculated the equilibrium lattice constant $a_{o}$, the bulk modulus ($B_{o}$), 
the associated total energy, and the electron and hole effective masses in different directions.

In calculating the lattice constant, we utilized a least square fit of our data to the  
Murnaghan's equation of state \cite{Murnaghan1995}. The lattice constant 
for the minimum total energy is the equilibrium one. 
The bulk modulus ($B_{o}$) is calculated at the equilibrium lattice constant.

\begin{figure}
\centering
%FIG. 3
 \includegraphics*[trim = 0mm 0mm 0mm 80mm,width=1\columnwidth,clip=true]{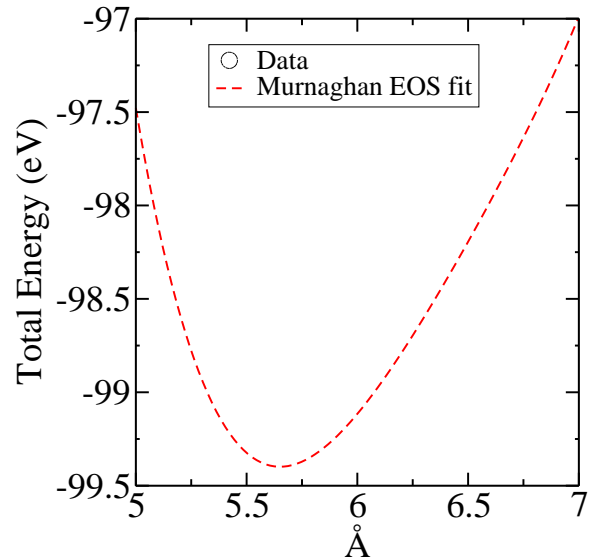}
\caption{(Color online). Calculated total energy per unit cell (E$_\mathrm{T}$) as a function of the lattice constant (\AA{}) 
of Ge, as obtained with the GGA BZW-EF optimal basis set. 
The calculated equilibrium lattice constant is 5.63 \AA{}, is little smaller than the room experimental value of 5.66 \AA{}, 
as expected.} 
\label{fig:ge_eos}
\end{figure}

The difference in energy between the GGA BZW-EF and LDA BZW-EF computational results is small 
$\sim$ 0.05 eV. Thus, the results reported here are those from GGA BZW-EF unless otherwise stated. 
The results of the electronic structure are given in 
Figs.~\ref{fig:ge_bnd} and \ref{fig:ge_pdos}. Total energy versus lattice constant data are as shown in 
Fig.~\ref{fig:ge_eos} while Table~\ref{table:ge_energy} shows the eigenvalue energies at various 
high symmetry points in the Brillouin zone. 

Our ab-initio, first principle results show that the fundamental gap of Ge is an indirect one with the maximum of 
the valence band ($VB_{max}$) occurring at $\Gamma$ and the minimum of the conduction band 
($CB_{min}$) at the $L$ point (cf. Figs. ~\ref{fig:ge_bnd} and ~\ref{fig:ge_pdos} and Table~\ref{table:ge_energy}). 
Our calculated indirect gap, using the equilibrium lattice 
constant, is 0.649 eV, while the indirect band gap value using the experimental lattice 
constant is 0.652 eV. Our calculated direct 
gap at the $\Gamma$ point is 0.857 eV (cf. Table~\ref{table:ge_energy}). Our calculated indirect 
band gap value using the LDA BZW-EF is 0.644 eV with a corresponding direct gap of 0.857 eV. The LDA and GGA potentials 
lead to the same direct gap, while the LDA indirect gap is 8 meV smaller than that for GGA. 
Our calculated values are in good agreement with experimental room temperature 
indirect band gap value of 0.664 eV and direct band gap value in the range 0.805 -- 0.895 eV \cite{Madelung2006,Adachi2005}. 
The accurate, calculated values for the indirect and direct gaps of Ge are a significant 
improvement over some previous results that employed methods (or approaches) beyond DFT 
\cite{PhysRevB.64.245204,PhysRevLett.89.126401,
PhysRevB.62.7121,PhysRevB.48.17791,PhysRevB.34.5390}.  
Our calculated spectra shown in Fig.~\ref{fig:ge_pdos}, especially 
for the valence states, are in good agreement with experiments \cite{PhysRevB.9.600,
PhysRevB.12.4405,PhysRevB.47.2130,PhysRevB.73.035217,PhysRevB.49.16523,PhysRevLett.29.1088}. 
The same can be said of our calculated 
bands, as shown in Fig.~\ref{fig:ge_bnd}, which are in good agreement with experiment 
\cite{PhysRevB.42.7429,PhysRevB.32.2326}. 
The occupied bandwith is calculated to be 12.56 eV, in very good agreement with room 
temperature values of 12.6 $\pm$ 0.3 \cite{Hellwege1969,PhysRevB.9.3473,PhysRevB.12.4405,PhysRevB.42.7429}. 

From Fig.~\ref{fig:ge_pdos}, in the 
DOS of valence states, we predict two weak shoulders at -0.31 and -0.67 eV, and a peak at -1.50 eV. This peak is followed by 
relatively broad ones at -1.90, -2.41, and -2.86 eV, before two sharp peaks at -4.06 and -7.41 eV. 
In the conduction bands, a small shoulder can be found at 0.71 eV followed by a peak at 2.10 eV. Other 
peaks can be seen at 3.07, 3.88, 4.44, and 
5.06 eV. The position of these critical points are in good agreement with experiments 
\cite{PhysRevB.9.3473,PhysRevB.42.7429,PhysRevB.47.2130,PhysRevB.44.13791,PhysRevB.9.600,PhysRevB.12.4405,
PhysRevB.73.035217,PhysRevB.49.16523,PhysRevB.81.035313,PhysRevB.33.2607,PhysRevLett.56.1187,PhysRevB.28.1965}. 
For instance, scanning tunneling spectroscopy measurements of \citet{PhysRevB.44.13791} reported peaks in the valence states at 
-0.47 $\pm$ 0.06 eV and -1.23 $\pm$ 0.030 eV while the inverse photoemission spectra of 
\citet{PhysRevLett.56.1187} reported peaks at 2.40 $\pm$ 0.10, 4.20 $\pm$ 0.10, and 5.50 $\pm$ 0.10 eV. We note that our 
spectra are not broadened as it is the case in (many) experiments; they may therefore have more features.

From Fig.~\ref{fig:ge_pdos}, significant contributions of 3$d$ states can be seen in the 
conduction bands and minimal contribution in the valence bands. 
Even though Ge is known to be an $sp$ material, the correct treatment of 3$d$ states 
and others as valence ones is critical for accurate results. Crystal symmetry gives rise to polarizations that 
are absent in isolated atom. In our work, only Ge (1$s$2$s$2$p$) were considered to be in the core. 
The importance of this view has been highlighted by the work of 
\citet{PhysRevLett.69.2955} and confirmed by those of \citet{PhysRevLett.89.126401} and of \citet{PhysRevB.48.17791}. 
It generally leads to a better description of the band structure energies, 
especially in obtaining the fundamental, indirect gap of Ge. Our computations with the 
3$d$ states treated as core electrons led to an indirect gap of 0.37 and 0.42 
for the LDA and GGA potentials, respectively. This goes to show the importance of the methodical search 
for the \textit{optimal basis set} as it is done in the BZW-EF. If we neglect the 3$d$ states in our 
computations, the gap closes making Ge a metal.

Effective masses are measures of the accuracy of a method as they depend very sensitively on the curvatures of the bands. 
They determine in part the transport properties, 
the Seebeck coefficient, and
the electrical conductivity of materials. The calculated 
electron effective masses at the bottom of the conduction band along 
the $\Gamma$ - $L$, $\Gamma$ - $X$, and $\Gamma$ - $K$ 
directions, are 0.043 - 0.051, 0.043 - 0.051, and 0.043 - 0.052 (all in units of the electron mass), 
respectively. These results are 
in good agreement with experimental value of 0.038 -- 0.083 \cite{Hellwege1969}. 

The total energy versus the lattice constant data are shown in Fig.~\ref{fig:ge_eos}. 
The data fit well to the Murnaghan equation of 
state (EOS) \cite{Murnaghan1995}. 
The calculated bulk modulus, $B_{o}$, is 80 GPa, in close agreement with 
experimental values of 75 -- 80 GPa \cite{Madelung2006,mcskimin:3312,mcskimin:988}. 
We also calculated the equilibrium lattice constant from the total energy minimization. The calculated 
equilibrium lattice constant is 5.63 \AA{}. The experimentally reported room temperature lattice constants are in the range 
of 5.62 to 5.66 \AA{}
\cite{Madelung2006,Wyckoff1963}. 

In summary, our results show that the electronic and related properties of Ge can be accurately described, with LDA/GGA 
potentials, by 
a careful search for a basis set that is verified to be complete for the description of the ground states.  

This work was funded in part by the the National Science Foundation 
[Award Nos. 0754821, EPS-1003897, NSF (2010-15)-RII-SUBR, and HRD-1002541], 
the US Department of Energy -- National Nuclear Security Administration (NNSA) 
(Award No. DE-NA0001861), NASA, through the Louisiana Space Consortium 
[LaSPACE, Award No. 5-27316], and by the Louisiana Optical Network Initiative (LONI) 
at Southern University and A\&M College in Baton Rouge (SUBR). 
C. E. Ekuma wishes to thank the Government of Ebonyi State, Nigeria.

\begin{widetext} \label{sup:supp}
 \begin{center}
 \textbf{SUPPLEMENTARY MATERIAL}
 \end{center}
Supplementary material with the detailed description of the BZW-EF method and the progress already made in 
tackling the band gap problems in materials related to this article can be found online
at
\end{widetext}

%\bibliography{Ge}

\end{document}

\title{[Supplementary Information]\\ Re-examining the electronic structure of germanium: A first-principle study}
\author{C. E. Ekuma}
\altaffiliation{corresponding email: cekuma1@lsu.edu}
\affiliation{Department of Physics \& Astronomy Louisiana State University,
Baton Rouge, LA 70803, USA}
\affiliation{Center for Computation and Technology, Louisiana State University, Baton Rouge, LA 70803, USA}

\author{M. Jarrell}
\affiliation{Department of Physics \& Astronomy Louisiana State University,
Baton Rouge, LA 70803, USA}
\affiliation{Center for Computation and Technology, Louisiana State University, Baton Rouge, LA 70803, USA}

\author{J. Moreno}
\affiliation{Department of Physics \& Astronomy Louisiana State University,
Baton Rouge, LA 70803, USA}
\affiliation{Center for Computation and Technology, Louisiana State University, Baton Rouge, LA 70803, USA}

\author{D. Bagayoko}
\affiliation{Department of Physics, Southern University and A \& M 
College, Baton Rouge, LA 70813, USA}

%\date{\today}

\maketitle

\section{DFT and Progress in band gap in materials}
Despite the great progress made possible by density functional theory (DFT), 
from 1964 to present, problems associated with obtaining theoretically the 
measured energy or band gaps, for finite and crystalline semiconductors, 
respectively, have persisted. Specifically, most DFT calculations, with 
emphasis on those utilizing local density approximation (LDA) and 
semi-local potentials, have led to semiconductor band gaps 
that are 30 -- 50\% smaller than their corresponding, measured values. 
Much effort has been deployed to find explanations of and remedies to 
this recalcitrant band gap problem. Perdew and Zunger 
\cite{PhysRevB.23.5048} introduced the self interaction correction 
(SIC) to local spin density (LSD) approximation calculations. 
While the exact functional for the ground state is self interaction 
free, these authors discussed corrections that appear to be needed 
for the description of finite systems, beginning with atoms, 
and of localized states in solids.  This self interaction is 
argued to contribute to the underestimation of the band gaps 
of insulators by DFT calculations \cite{PhysRevB.23.5048}. Consequently, 
self interaction corrections (SIC) are expected to improve the 
agreement between calculated band gaps and measured ones, in 
addition to improving binding energies and bringing orbital 
energies closer to removal energies \cite{PhysRevB.23.5048,PhysRevB.26.5445}. 
While self interaction corrections have led to some improvements 
in band gap calculations, they have not totally resolved the problem. 
Applications of SIC have mostly overestimated the band gap of semiconductors 
\cite{PhysRevB.52.R14316,PhysRevB.54.5495}. According to Cohen \etal \cite{Cohen2008}, 
self interaction is well-defined only for one-electron systems.

According to the literature, a major source of the theoretical underestimation 
of band gaps consists of the derivative discontinuity of the exchange correlation 
energy, Exc \cite{PhysRevLett.51.1888,PhysRevLett.49.1691,PhysRevLett.51.1884,PhysRevB.32.3883}. 
Perdew \etal \cite{PhysRevLett.49.1691}, following a thought experiment on a 
diatomic molecule, established the existence of a derivative 
discontinuity of the exchange correlation energy, i.e., a 
discontinuity in the exchange correlation potential, V$_{xc}$. 
Perdew and Levy \cite{PhysRevLett.51.1884} generalized this discontinuity 
to the case of semiconductors. They showed that the exchange 
correlation potential may jump by the discontinuity, $\Delta_{xc}$, 
when the number of electrons in the system under study 
increases by one.  Band gaps calculated with a local 
density approximation (LDA) potentials, according to 
their findings, are to be augmented by this discontinuity 
in order to reproduce the corresponding, measured values. 
The authors suggested, without claiming to have a proof of it, 
that this discontinuity is a non zero (and positive) in 
real semiconductors and insulators.  Sham and Schl\"{u}ter 
\cite{PhysRevLett.51.1888} also found a derivative discontinuity of 
E$_{xc}$ in insulators. These authors, however, asserted 
that their work does not show whether or not this discontinuity is 
non zero in real insulators.   Subsequent work by 
Sham and Schl\"{u}ter \cite{PhysRevB.32.3883} derived the discontinuity 
of the functional derivative of E$_{xc}$ in insulators by considering 
an increase of the number of electrons by one. Cautiously, 
these authors concluded that the discrepancy between 
calculated and measured band gap is a measure of 
the discontinuity  $\Delta_{xc}$ - given the results from several
calculations -- if the employed LDA potentials are assumed 
to be good approximations. The description of our method 
below indicates the strong possibility that some current 
LDA and GGA potentials may be very good approximations. 

Despite its popular use to explain the disagreement between 
calculated and measured band gaps, the above discontinuity 
has not yet been established to be non zero in real 
semiconductors or insulators.  Further, Sham and Schl\"{u}ter 
\cite{PhysRevB.32.3883} underscored the fact that, in principle, DFT 
and Kohn Sham LDA hold only if the number of particle is kept constant. 
The question could arise whether or not the discontinuity, derived by 
considering a change of the number of particle, is strictly applicable 
to DFT or LDA calculations. From the preceding, it has not yet been 
established that DFT or LDA calculations cannot obtain the correct band gaps, 
despite the fact that presently known LDA potentials do not have a 
discontinuity and that most of the numerous,  
previous ab-initio DFT and LDA calculations did not. 

Another presumed contributor to the band gap underestimation by theory stems from the use of 
local (LDA) and semi-local (GGA) potentials. 
The question naturally arises as to what extent the local and semi-local 
potentials fail to capture key feature of the exact one. We are aware of 
no definitive answer, given that the exact one is not known. We would have 
had to delve into this matter further if we were dealing with molecules or 
their dissociation. The solid state systems of interest to us, to judge by 
previous results obtained with our method \cite{Bagayoko2004,Bagayoko2008}, 
possible errors due to the use of local and semi-local potentials appear to be very small. 

There exist several approaches that have been introduced to address the band 
gap problem. Review articles and books are the best sources for discussing 
these approaches and for examples of the many DFT calculations that 
led to band gaps much smaller than their corresponding, experimental counterparts. 
In contrast, a summary of results from BZW LDA calculations for over 
10 materials show agreement between theory and experiment. 
Illustrative examples of discrepancies between theory and experiment follow. 
The case of Ge is summarized in this article. 
Some previous LDA, GGA, and GW calculations did not yield the measured band gap, from first principle. 
A table provided by Ekuma and Bagayoko \cite{Ekumab2011} shows a multitude of DFT 
calculations with vastly different band gaps for titanium dioxide.  
With the computational method described here, Ekuma and Bagayoko 
obtained the measured, direct gap and predicted an indirect one.  
For elemental silicon, Gr\"{u}ning \etal \cite{Myrta2006} 
reported an LDA band gap of 0.7 eV, much smaller than the 1.25 eV 
they reported as the measured value.  These authors also performed 
calculations with the exact exchange (EXX), EXX plus LDA, EXX plus 
the random phase approximation (RPA). 
The last approach or scheme yielded 0.6 eV, a gap smaller than the above LDA gap, 
while the first two led to 1.5 and 1.6 eV respectively, values 
much larger than the experimental one. With the
original version of our method,  Zhao et al. utilized an LDA potential to 
obtain a gap of 1.02 for Si, much closer to the experimental one. 
Generalized gradient approximation (GGA) calculations have led to 
improvements of calculated properties of materials, including lattice parameters. 
Specifically, Hao \etal \cite{PhysRevB.85.014111} reported 
revised Tao-Perdew-Staroverov-Scuseria (revTPSS) meta-GGA 
calculated lattice parameters that are in agreement with 
experiment following a zero-point phonon correction, 
for over 50 materials. Despite this very significant success, 
most GGA and meta-GGA calculations, including the previous 
ones discussed here for Ge, have not produced band gaps in agreement with experiment. 

From the above summary, historical overview of the band 
gap problem, it appears that the scientific community believes 
that the derivative discontinuity of the exchange correlation energy 
is the main source of the disagreement between DFT calculated energy 
and band gaps and their corresponding, measured ones.  This belief 
led to the development of several schemes aimed at resolving the band 
gap problem. Except for the few, most of these schemes are ad hoc as 
they include adjustable parameters that vary with the material under 
study. The continuing growth in the number of these schemes seems 
be a problem in itself, the ad hoc nature of most of them does not 
lend itself to predictive capabilities from first principle, the aim 
of theory to inform and to guide experiment. The only exception to 
the above trend consists of the work of our group. This work has 
not yet gotten the attention of the community at large, presumably 
due to the strength of the above belief, on the one hand, and the 
preponderance of results that are explained with the discontinuity, on the other hand.  
As we previously noted \cite{Bagayoko2005}, 
the situation resembles that of the Ptolemaic model of the solar system 
where epicycles were continually introduced to explain its disagreement 
with observations.  The quintessential point in support of the our 
method, described below, is the following: For all DFT calculations 
of energy bands, the \textit{minima of the occupied energies, which add up to 
yield the ground state energy of the electron system, are obtained from 
the theory if the ``correct'' ground state charge density is utilized, 
subject to the constraint that the number of particle is kept constant} 
\cite{PhysRev.136.B864,PhysRev.140.A1133}. 

Most of the previous DFT calculations, including those with GGA 
and LDA potentials, have consisted of judicious selecting large 
basis set and of performing iterations to obtain self consistent 
eigenvalues of the Kohn-Sham type equation. It is assumed that 
the single basis set in question leads to the correct representation 
of the electronic cloud in the system under study, a system that can be 
drastically different from an atomic or ionic one. \textit{In particular, as we 
recently pointed out, polarization ($p$, $d$, and $f$ orbitals) has primacy over 
spherical symmetry ($s$ orbital) for systems varying from diatomic 
molecules to solids.}  Hence, utilizing basis sets derived from 
calculations of properties of atoms for the study of 
solids is potentially problematic. Indeed, the angular symmetries 
in these systems are vastly different
from those for atoms and ions.  The need for 
the method described below becomes apparent with 
the realization that, irrespective of the degree of 
convergence of the iterations, a single trial basis 
set that has a major symmetry inadequacy for the 
description of the system is not going to lead to physically valid DFT 
eigenvalues as the implacable condition of using the 
``correct'' ground state density will not be met. 

\section{Description of the Bagayoko-Zhao-Williams-Ekuma-Franklin Method}
The original version of our method, named after Bagayoko, 
Zhao, and Williams (BZW) was introduced in 1998 \cite{Bagayoko1998}
and further explained in 1999 
\cite{Zhao1999}. The method consists of 
implementing the linear combination of atomic orbitals (LCAO) 
formalism by methodically searching for the smallest basis 
set, called the optimal basis set, that leads to the minima 
of the occupied energies. This search begins with a 
deliberately small basis set that is not smaller than 
the minimum basis set, i.e., the smallest one needed 
to account for all the electrons in the system. 
The self consistent calculation with this basis set 
is followed by another whose basis set uses the previous 
one plus one additional orbital. The occupied energies 
from the two calculations are compared numerically and 
graphically. In the more than 20 systems we have studied, 
these occupied energies from these first two calculations 
have been different, with those of calculation II generally 
lower than the corresponding one from Calculation I. 
Calculation III is then carried out, using the basis set 
in Calculation II plus an additional orbital. 
The occupied, self consistent energies from Calculations 
II and III are also compared. This process of augmenting 
a basis set, performing new self-consistent calculations, and comparing 
its results with those of the one immediately preceding it, 
continues until a calculation is found, say N, to have 
exactly the same occupied energies as the one immediately 
following it.   This perfect superposition establishes that 
the minima of the occupied energies have been reached and 
that the corresponding basis set give the best representation 
of the ground state charge density of the system. 

Before elaborating further on the physical content of 
the method, we note that adding an orbital means 
increasing the size of the basis set (and hence the 
dimension of the Hamiltonian) by 2, 6, 10, or 14 depending 
on the $s$, $p$, $d$, or $f$ character, respectively, of the orbital 
in question.  In the original BZW, we added orbitals in the order 
of their energies resulting from the atomic calculations, i.e., 
the orbitals corresponding to the lowest laying excited, atomic 
state were successively added.  As apparent from our earlier results, 
the BZW method practically led to the measured 
band gap of semiconductors we studied. Further, our predictions of 
the band gaps and other properties for cubic Si$_\mathrm{3}$N$_\mathrm{4}$ \cite{Bagayoko2001} 
and cubic InN \cite{Bagayoko2004} were totally confirmed by 
experiment \cite{PhysRevB.65.161202,Egdell2003,Sch2006}. 
Following the works of Ekuma and Franklin \cite{Ekuma2011,ekuma:012189,Ekumab2011,ibid2011}, 
we realized that valence electrons in multi-atomic systems simply 
do not follow the symmetry landscape that prevail for isolated atoms or ions.  
The aim is to obtain a better representation of the electronic cloud 
(ground state charge density) of the system
under study. Hence, in most of our subsequent calculations, 
for a given principal quantum number, we add $p$, $d$, and $f$ 
(if applicable) orbitals before the $s$ orbital for that principal quantum number. 
This counter-intuitive ordering, for isolated atoms, is simply 
needed for `some' multi-atomic systems.  The initials of Ekuma and 
Franklin (EF) are added to the name of the enhanced method [BZW-EF] 
in recognition of their extensive calculations whose results led 
Bagayoko to see the necessity for this new order. While the 
BZW method led to band gaps that were sometimes smaller 
by 0.1 -- 0.3 eV, insignificant differences between 
BZW-EF calculated gaps and corresponding experiment 
ones are in the second decimal place for the 
systems studied to date. For ZnO \cite{Franklin2013729}, the BZW-EF method 
led to an upper valence band width more than 
1.0 eV larger than was obtained with the BZW, 
with a significant improvement in agreement with experiment. 

The origin of the changes in the band structure 
and the band gap when the basis set increases toward 
the optimal one consists of the progressively better 
representation of the ground state charge density. As 
per the derivation of DFT, the minima of the occupied 
energies are obtained if the ``correct'' charge 
density for the ground state is employed. These changes 
are due to physical interactions, given that the charge 
density, the potential and hence the Hamiltonian change 
from one calculation to the next.  We should underscore 
here that while our focus is on occupied energies 
(i.e., DFT is a fundamentally ground state theory), 
when these energies reach their minima, so do the low 
laying unoccupied energies, up to 9 -- 10 eV for the materials 
studied to date with the BZW-EF method. For the many 
systems with the BZW, most low 
laying unoccupied energies also converged up to 
5 -- 6 eV, as was the case for wurtzite indium nitride \cite{Jin2007}. 
For a few materials, this convergence 
of the lowest unoccupied energies was not achieved 
with the original BZW method. For metals, as shown by Zhao \etal \cite{Zhao1999}, the low-laying 
unoccupied energies converge when the occupied ones do, 
due to the fact that at least one band crosses the Fermi level. 
This fact may partly explain the early successes of 
DFT in describing metals as compared to semiconductors. 

The description of what occurs when basis 
sets much larger than the optimal ones are 
employed will complete the description of the 
BZW-EF method.  We first recall that the basis set 
immediately following the optimal one leads to the 
occupied energies obtained with the optimal one and 
to the same unoccupied energies up to 9 -- 10 eV. So, 
by much larger basis sets, we mean the ones that are 
larger than the basis sets immediately following the 
optimal ones. Earlier works by Bagayoko and Co-workers \cite{Zhao1999,Bagayoko1998} verified 
that basis sets larger than the optimal one do not change 
the charge density, the potential, and the Hamiltonian, nor 
do they change the occupied energies. In the absence of changes 
in the Hamiltonian, i.e., the physics of the study, the additional 
lowering of unoccupied energies with these much larger basis set 
cannot be ascribed to DFT. However, the Rayleigh theorem provides 
an explanation of the unphysical lowering
of these energies. The theorem states that when the 
same eigenvalue equation is solved with two basis sets 
of different sizes, such that the larger one includes 
the smaller one, then the eigenvalue obtained with the 
larger basis set are lower than or equal to their corresponding one 
obtained with the smaller basis set.  

Clearly, after the optimal basis set is reached, and that 
the occupied energies are no longer changed from their values 
obtained with the optimal one, the lowering of unoccupied energies 
can be ascribed to a mathematical artifact that is the manifestation 
of the above theorem. We therefore contend that this extra lowering, a 
variational basis set effect \cite{Zhao1999,Bagayoko1998}, 
is a major source of discrepancies between many previous DFT calculations  
and between these calculations and experiment, as far as band gaps are concerned. 
\textit{The lowest laying conduction bands, with full physical meaning when they result 
from the optimal basis set, partly lose their physical content due to the above 
effect.} It is important to note that this is the case for any basis set that is 
not the optimal one, whether it is smaller or larger than the optimal, or 
simply lacks orbital or angular features of the optimal one. For basis sets 
that do not totally include their corresponding optimal ones, even the 
occupied energies are not totally DFT results.  The preceding lines in this 
paragraph point to the great difficulty in obtaining physically meaning DFT 
occupied and low laying unoccupied energies with a single 
trial basis set, irrespective of the degree of convergence of the 
applicable iterations for self consistency.

%\bibliography{Ge}